%% file: main.tex
\documentclass[sigconf]{acmart}
\AtBeginDocument{%
  }

\setcopyright{acmlicensed}
\copyrightyear{2018}
\acmYear{2018}
\acmDOI{XXXXXXX.XXXXXXX}
\acmConference[Conference acronym 'XX]{Make sure to enter the correct
  conference title from your rights confirmation email}{June 03--05,
  2018}{Woodstock, NY}
\acmISBN{978-1-4503-XXXX-X/2018/06}

\newcommand{\rqone}{RQ1: Which retrieval approach configurations yield the highest response quality in LLM-generated answers?}
\newcommand{\rqtwo}{RQ2: How well does adaptive HyDE retrieval perform on novel questions outside the training corpus?}
\newcommand{\rqthree}{RQ3: How does our proposed RAG pipeline perform across different LLMs?}

\usepackage{enumitem}
\usepackage{makecell}

\usepackage{multirow}
\usepackage{tabularray}
\usepackage{siunitx}
\usepackage{tcolorbox}
\tcbuselibrary{skins,breakable}
\usepackage{fontawesome}
\usepackage{listings}
\usepackage{xcolor}  

\definecolor{mygreen}{HTML}{4D9900}

\tcbset{
  promptstyle/.style={
    enhanced, sharp corners, 
    colback=blue!4!white, colframe=blue!60!black,
    boxsep=1mm, left=1mm, right=1mm, top=0.8mm, bottom=0.8mm,
    fonttitle=\scriptsize\bfseries,
    fontupper=\scriptsize}
}

\usepackage{svg}
\begin{document}

\title{Never Come Up Empty: Adaptive HyDE Retrieval for Improving LLM Developer Support}


\author{Fangjian Lei}
\email{fangjian.lei@queensu.ca}
\authornotemark
\affiliation{%
  \institution{Queen's University}
  \city{Kingston}
  \state{Ontario}
  \country{Canada}
}
\author{Mariam El Mezouar}
\email{mariam.el-mezouar@rmc.ca}

\affiliation{%
  \institution{Royal Military College of Canada}
  \city{Kingston}
  \state{Ontario}
  \country{Canada}
}
\author{Shayan Noei}
\email{s.noei@queensu.ca}
\affiliation{%
  \institution{Queen's University}
  \city{Kingston}
  \state{Ontario}
  \country{Canada}
}
\author{Ying Zou}
\email{ying.zou@queensu.ca}
\affiliation{%
  \institution{Queen's University}
  \city{Kingston}
  \state{Ontario}
  \country{Canada}
}


\renewcommand{\shortauthors}{Trovato et al.}

\include{sections/0_abstract}



\renewcommand\footnotetextcopyrightpermission[1]{}
\setcopyright{none} 
\settopmatter{printacmref=false} 

\maketitle

\input{sections/1_intro}

\input {sections/3_overall_approach}

\input {sections/4_experiment}
\input {sections/5_rq}

\input {sections/6_discussion}

\input {sections/7_threats}
\input {sections/8_related_work}

\input {sections/9_conclusion}

\bibliographystyle{ACM-Reference-Format}
\bibliography{references}

\appendix

\end{document}

%% file: sections/0_abstract.tex
\begin{abstract} 
Large Language Models (LLMs) have shown promise in assisting developers with code-related questions; however, LLMs carry the risk of generating unreliable answers. To address this, Retrieval-Augmented Generation (RAG) has been proposed to reduce the unreliability (i.e., hallucinations) of LLMs. However, designing effective pipelines remains challenging due to numerous design choices. In this paper, we construct a retrieval corpus of over 3 million Java and Python related Stack Overflow posts with accepted answers, and explore various RAG pipeline designs to answer developer questions, evaluating their effectiveness in generating accurate and reliable responses. More specifically, we (1) design and evaluate 7 different RAG pipelines and 63 pipeline variants to answer questions that have historically similar matches, and (2) address new questions without any close prior matches by automatically lowering the similarity threshold during retrieval, thereby increasing the chance of finding partially relevant context and improving coverage for unseen cases. We find that implementing a RAG pipeline combining hypothetical-documentation-embedding (HyDE) with the full-answer context performs best in retrieving and answering similar content for Stack Overflow questions. Finally, we apply our optimal RAG pipeline to 4 open-source LLMs and compare the results to their zero-shot performance. Our findings show that RAG with our optimal RAG pipeline consistently outperforms zero-shot baselines across models, achieving higher scores for helpfulness, correctness, and detail with LLM-as-a-judge. These findings demonstrate that our optimal RAG pipelines robustly enhance answer quality for a wide range of developer queries—including both previously seen and novel questions—across different LLMs.
\end{abstract}

%% file: sections/1_intro.tex
\section{Introduction}
\label{sec:introduction}
Programmers often rely on online resources for a wide range of development tasks, such as API usage, bug fixing, and understanding of code or programming concepts~\cite{rao2020analyzing, skripchuk2023analysis, xia2017developers}. A significant portion of these help-seeking activities involves regular interaction with community-driven Q\&A platforms like Stack Overflow (SO)~\cite{rahman2018evaluating, vasilescu2013stackoverflow, xia2017developers}. Recently, the emergence of Large Language Models (LLMs) has begun to reshape how developers search for assistance in programming activities that developers increasingly prefer using conversational LLMs over traditional search methods like forums or search engines for programming assistance\cite{ross2023programmer}. Open-source LLMs such as LLaMA family~\cite{touvron2023llama}, have shown strong performance in code understanding and generation tasks, gaining increasing attention among software practitioners and researchers. These models offer the potential to serve as an alternative to traditional search on Q\&A platforms, enabling more conversational and context-aware support during programming tasks.

Despite the rising popularity of Large Language Models (LLMs) used for information seeking, there are growing concerns about the reliability and correctness of generated content commonly referred to as hallucination. Previous studies have shown that LLMs can learn incorrect information during training and later reproduce or even amplify these errors in the generated outputs~\cite{bender2021dangers, gamage2022deepfakes, gehman2020realtoxicityprompts}. LLMs are also capable of producing fabricated content that mimics truthful responses, which can be difficult to detect, especially for users without domain expertise~\cite{bommasani2021opportunities, cao2021knowledgeable}. 

To mitigate hallucination, Retrieval-Augmented Generation (RAG) has emerged as a promising solution and has shown strong potential in improving the quality of responses generated by LLMs when a knowledge base with similar context is available for reference. RAG enhances LLMs by incorporating external knowledge retrieved from a document corpus into the generation process~\cite{lewis2020rag}. However, the effectiveness of RAG is highly dependent on the retriever's ability to identify relevant information. When the input question is novel or falls outside the scope of the retrieval corpus, existing RAG retrievers often struggle to extract useful content \cite{gilson2024enhancing,cheng2025survey}. Since existing RAG systems rely solely on the input question, they may fail to retrieve semantically relevant documents in such cases. The final answer depends largely on the LLM’s pre-trained knowledge and its ability to generalize. This limitation highlights the need for methods that generate informative answers even when relevant content cannot be retrieved, ensuring consistent performance across diverse questions.

\begin{figure*}[t]
    \centering
\includegraphics[width=1.0\textwidth]{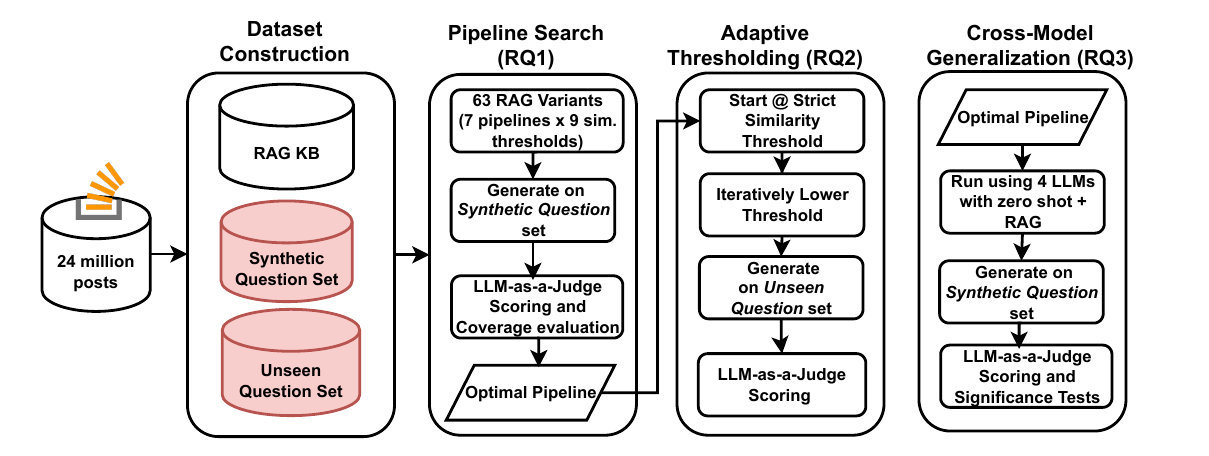}
    \caption{Experimental Workflow. \textit{\textbf{RAG KB}} = Stack Overflow Knowledge Base (3.4 M accepted-answer documents). \textit{\textbf{Synthetic Question Set}}: 385 questions auto-generated from the KB (seen). \textit{\textbf{Unseen Question Set}}: 5,510 new Stack Overflow questions posted after the KB snapshot (unseen).
}
    \label{fig:workflow}
\end{figure*}
In this paper, we aim to address the limitations of existing RAG approaches, which struggle with vague questions and often fail on novel questions. 
More specifically, we explore two implementations of Retrieval-Augmented Generation (RAG): (1) a question-based approach that searches the knowledge base using the original question, and (2) the Hypothetical Document Embedding (HyDE) approach \cite{gao2023precise}, which first generates a hypothetical answer to improve the relevance of retrieved content. The two RAG implementations are further characterized by three key design dimensions: the first dimension, \textbf{retrieval target}, determines whether content is retrieved directly from accepted answers or indirectly via similar questions. The second, \textbf{content granularity}, specifies whether the system retrieves full answers for broader context or individual sentences for more precise and relevant information. The third, \textbf{similarity threshold}, sets the semantic similarity score between the input and retrieved content, directly influencing the amount and quality of context for generation. These three design dimensions directly affect the amount and quality of context extracted from RAG knowledge base. Therefore, we conduct the experiments by systematically varying the dimensions to assess how different pipelines affect the quality of generated answers using LLMs based on RAG and identify the best-performing RAG pipeline that can extract relevant content for enhanced answer quality. 

\noindent In this paper, we aim to answer the following research questions: 

\textbf{\rqone} 
To determine how different design dimensions impact the effectiveness of RAG, we systematically evaluate 7 RAG pipelines and 63 pipeline variants that vary in retrieval target, content granularity, and similarity threshold. We assess each pipeline in terms of both answer quality and retrieval coverage on the Synthetic Question Set. Our results show that the hypothetical-answer-based pipeline (HB1), which retrieves from full answers in the knowledge base, consistently achieves the best trade-off between high response quality and broad coverage. This pipeline is selected as the optimal pipeline for further research questions.

\smallskip
\textbf{\rqtwo} 
Developers often pose novel questions that lack closely related content in the knowledge base, which limits the effectiveness of standard RAG methods. To address this, we extend our approach by dynamically decreasing the similarity threshold for each question until relevant context is retrieved. Evaluated on an unseen question set, dynamically decreasing the similarity threshold enables full coverage, ensuring every question receives relevant contextual from RAG. Results show that our method significantly improves answer quality over original Stack Overflow answers, with statistical analysis confirming the effectiveness of dynamic thresholding for unseen cases.

\smallskip
\textbf{\rqthree} 
Given the diversity of available LLMs, it is important to understand whether our optimal RAG pipeline offers consistent benefits across models. We apply the pipeline to several open-source LLMs and compare its performance to standard zero-shot prompting. Our findings reveal that our optimal RAG pipeline robustly improves or matches answer quality across different models, demonstrating strong generalization and practical value for a variety of LLM-based applications.


Our contributions are as follows:
\begin{itemize}[leftmargin=7pt,itemindent=0pt,labelsep=0.5em] 
    \item We present RAG frameworks for answering Java and Python questions, using Stack Overflow as a retrieval base and open-source LLMs for generating answers.
    \item We evaluate RAG implementations and propose a HyDE approach to improve answer retrieval performance on both seen and unseen questions.
    \item We provide an evaluation of multiple LLMs performance on developer questions in both matched (similar) and unmatched (unseen) scenarios.
    \item We release our dataset and pipeline to support future research in RAG-based methods for software engineering questions. Our replication package is available at~\cite{icse2026_stackoverflow_rag}.
\end{itemize}

The remainder of this paper is organized as follows. Section \ref{sec:methodology} details the proposed approach. Section \ref{sec:rqs} presents the research questions. Section \ref{sec:implications} is the discussion section. Section \ref{sec:threats_to_validity} addresses the threats to validity. Section \ref{sec:related_work} presents previous related work. Finally, Section \ref{sec:conclusion} concludes the paper and outlines future work.

%% file: sections/3_overall_approach.tex

\begin{figure*}[h]
    \centering
\includegraphics[width=1.0\textwidth]{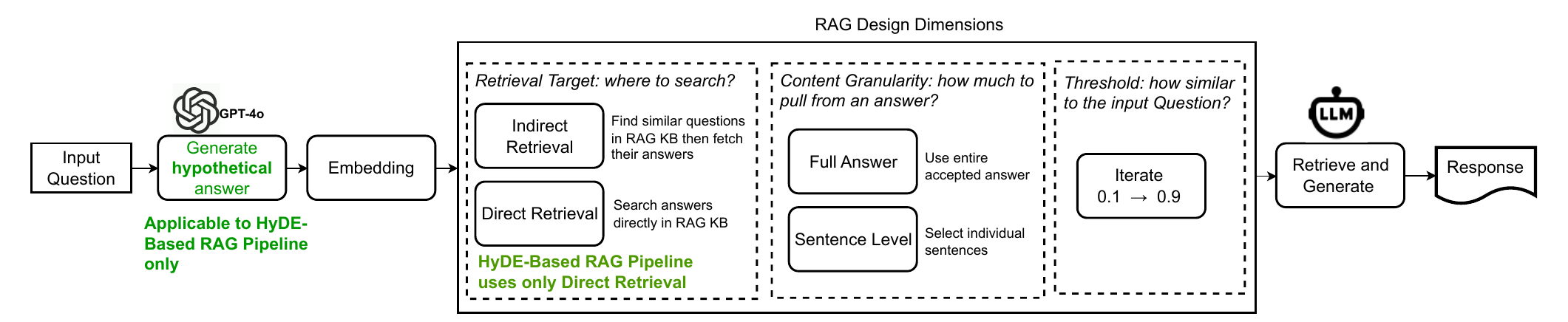}
    \caption{An Overview of Question-Based and HyDE-Based RAG pipeline. HyDE specific design elements shown in \textcolor{mygreen}{green}.}
    \label{fig:rag_implementation}
\end{figure*}

\section{Methodology}

\label{sec:methodology}

Figure~\ref{fig:workflow} outlines our experimental workflow. We first construct a \textbf{RAG Knowledge Base} of 3.4 million Java and Python Stack Overflow posts with accepted answers. Then we create two evaluation sets: (1) \textbf{Synthetic Question Set} of 385 generated questions to represent "seen" cases, and (2) \textbf{Unseen Question Set} of 5,510 posts made after May 2023 to simulate novel queries. Our evaluation proceeds in three stages: (RQ1) identifying the optimal RAG pipeline by varying design choices and assessing both answer quality using the LLM-as-a-Judge framework and retrieval coverage on the \textbf{Synthetic Question Set}; (RQ2) introducing adaptive thresholding to improve retrieval for novel questions and evaluating its impact on the \textbf{Unseen Question Set}; and (RQ3) testing the generalizability of our optimal pipeline across multiple open-source LLMs using \textbf{Synthetic Question Set}.

\subsection{Dataset Construction}

To support domain-specific retrieval, we build a \textbf{RAG Knowledge Base} from the official Stack Overflow data dump,\footnote{\url{https://archive.org/details/stackexchange}} covering posts from January 2008 to December 2024. We focus on posts tagged with \texttt{[java]} or \texttt{[python]} because of the sustained relevance in both educational and industrial contexts \cite{cutting2021comparative}. The dataset construction process involves the following steps:\\
\noindent\textbf{Filtering:} We retain only non-duplicate questions with an accepted answer to ensure reliable question–answer pairs for retrieval and evaluation. After filtering the raw Stack Overflow data, we obtain a total of 3,428,217 posts tagged with either \texttt{[java]} or \texttt{[python]} that include an accepted answer, which we use to construct our RAG knowledge base.
    
\noindent\textbf{Cleaning:} We remove HTML tags and markdown syntax to produce clean plain-text content suitable for indexing and retrieval. In addition, we split the accepted answers into individual sentences to support sentence-level retrieval. We also retain the complete accepted answers to support both sentence- and answer-level content granularity.
    
\noindent\textbf{Splitting:} We reserve posts from May 2023 to December 2024 as the evaluation set to simulate future or unseen queries, while using earlier posts for retrieval to mitigate data leakage. To evaluate the performance of the RAG pipelines, we construct two distinct question sets: \textbf{Synthetic Question Set} and \textbf{Unseen Question Set}. To evaluate the performance of various RAG pipelines, we sample and synthesize 385 questions from the knowledge base to construct \textbf{Synthetic Question Set}. Each question is created by sampling a question from the Stack Overflow dataset, then prompting GPT-4o\footnote{\label{openai}https://platform.openai.com/docs/models} to generate a similar question in order to reduce the data leakage issue. The selected sample size ensures a 95\% confidence level with a 5\% margin of error \cite{junk1999confidence}. This synthetic set is used as the main evaluation for assessing answer quality in RQ1 and RQ3. To test our approach on new and previously unseen questions (RQ2), we construct an \textbf{Unseen Question Set} containing 5,510 questions posted between May 2023 and December 2024. All questions in this set are posted after the release of \textit{LLaMA-3.1-8B-Instruct}\cite{meta2024llama3}, the model used for response generation. We begin by extracting all Stack Overflow posts tagged with either \texttt{[java]} or \texttt{[python]} from this time period and remove any posts whose question titles also appear in the RAG Knowledge Base. This yields a total of 11,771 questions—2,755 related to Java and 9,016 related to Python. To balance the evaluation set, we randomly sample 2,755 Python questions and combine them with all 2,755 Java questions, resulting in a final set of 5,510 testing cases. None of these posts appear in the RAG Knowledge Base, ensuring a true unseen scenario for evaluation.


\subsection{LLM-as-a-Judge Scoring Framework}
\label{sec:evaluation}
We adopt the \textit{LLM-as-a-Judge} mechanism to score every generated answer based on \textbf{helpfulness} (whether the response addresses the user query), \textbf{technical correctness} (technical accuracy), and \textbf{level of
detail} (completeness and depth). We feed the LLM-as-a-Judge prompts to GPT-4o, to produce a single composite score per answer on a 1–10 scale (based on the three criteria). We show the specific prompts used throughout the rest of the paper.
To verify that automatic scores reasonably align with human judgment, two independent annotators manually rate a sample of \textbf{54} answer pairs (chosen for a 90\% confidence level and a 10\% margin of error). Both annotators are graduate students in computer science and each has over 10 years of experience with Python, Java, and software development. For comparison purposes, we map the LLM scores to binary labels   
(\texttt{2–5}\,$\rightarrow$\,0, \texttt{6–9}\,$\rightarrow$\,1) \footnote{No answers in the sample received a raw score of 0, 1, or 10.}.
Cohen’s~$\kappa$ indicates \emph{moderate} agreement between the human annotators and LLM labels, which we consider acceptable for this evaluation.

\input{tables/rag_pipeline}

\subsection{Identifying Optimal RAG Pipelines}
RAG is used to enhance LLM responses by incorporating relevant content retrieved from a Stack Overflow Knowledge Base consisting of \textit{Java} and \textit{Python} posts with accepted answers. We introduce two main RAG implementations as shown in Figure \ref{fig:rag_implementation}. Based on the two main RAG implementations, we outline multiple pipeline variants with different component combinations.

\subsubsection{Two Base Implementations}.
\\
\noindent\textbf{Question-Based RAG:} This implementation directly uses the original input question to initiate retrieval. The question is first embedded using a sentence embedding model and then matched against a pre-processed Stack Overflow corpus to retrieve the most relevant content. The retrieved content is subsequently provided as context to a LLM, which in turn generates the final response.
    
\noindent\textbf{HyDE-Based RAG:} This implementation addresses the possible limitations of directly using the original question for retrieval. For example, although embedding models capture semantic similarity, short or vague questions are difficult to retrieve the most relevant answers, whereas a pseudo-answer provides a richer and more aligned representation for matching. To mitigate this, HyDE first generates a hypothetical answer using an LLM. This pseudo-answer tends to be more informative and semantically aligned with the expected answer format, making it a more effective query \cite{gao2023precise}. The pseudo-answer is then embedded and used to retrieve relevant content from the Stack Overflow corpus, which is passed to another LLM to generate the final response.

\subsubsection{Design Dimensions}

To improve RAG performance, we design multiple pipeline variants by systematically varying key design dimensions. As described below, we explore changes in content granularity, retrieval target. In Figure~\ref{fig:rag_implementation}, the dot-dashed boxes indicate the pipeline elements that can be adjusted or selected for each design dimension.

\noindent\textbf{Content Granularity:} Retrieved content can be either \textit{full accepted answers} or \textit{individual sentences}. Full-answer retrieval preserves coherence and broader context, but often includes irrelevant or redundant information \cite{nadi2020essential}. In contrast, sentence-level retrieval enables more precise selection of relevant content by filtering out unrelated or redundant text. Since not every sentence in an accepted answer contributes directly to answering the question, selecting at the sentence level allows us to incorporate multiple useful sentences from different answers. This increases coverage by capturing relevant information even from answers that do not directly or fully address the question.

\noindent\textbf{Retrieval Target:} This design dimension determines the initial source of retrieval within the Stack Overflow corpus. We explore two primary strategies: (1) indirect retrieval, which identifies questions that are similar to the input question, and then extracts their associated accepted answers or answer sentences; and (2) direct retrieval, by querying the answer corpus directly—either at the sentence or full-answer level. The first strategy benefits from the fact that semantically similar questions often have relevant answers, allowing it to retrieve useful content even if the input question is worded differently from questions in the Stack Overflow corpus.

\noindent\textbf{Similarity Threshold:}
A key configuration in both implementations is the choice of similarity threshold, which controls the trade-off between retrieving highly relevant but fewer results (using a high threshold) and achieving broader coverage at the cost of including less relevant content (using a lower threshold). 
In \textbf{Question-Based RAG}, the input question is first embedded into a vector representation. To support embedding computation, we employ the \textit{all-mpnet-base-v2} \cite{reimers2020making} model to embed input questions, accepted answers, and hypothetical content. \textit{all-mpnet-base-v2} is chosen for its strong performance on semantic similarity and retrieval tasks across diverse domains \cite{jayanthi2021evaluating}. The system then computes the cosine similarity \cite{gomaa2013survey} between the input question embedding and the embeddings of contents in the \textbf{RAG Knowledge Base}, such as question titles or answer sentences depending on the different pipeline dimensions design. Only candidates with a similarity score above the specified threshold are selected as relevant context for the next stage of generation. In \textbf{HyDE-Based RAG}, instead of using the original question, we first generate a hypothetical answer based on the question by using the GPT-4o. The hypothetical answer is then embedded, and its vector is compared to the embeddings of potential content (either full answers or individual answer sentences depending on the different pipeline dimensions design) in the \textbf{RAG Knowledge Base}. And, only those candidates whose similarity score exceeds the similarity threshold are extracted as relevant context.

\vspace{-1em}
\subsubsection{Pipeline Combinations}
Each pipeline is defined by a specific combination of \textbf{Retrieval Target} and \textbf{Content Granularity}. By varying the \textbf{similarity threshold} from 0.1 to 0.9, we produce nine distinct variants for each pipeline. In total, Table~\ref{tab:rag_variants} provides an overview of all 7 RAG pipelines and their 63 (7x9) unique pipeline variants, along with the design dimension descriptions.

\input{tables/similarity_threshold}

\subsection{Introducing the Adaptive Thresholding Strategy}
When developer questions do not have close matches in the knowledge base, retrieval-augmented generation (RAG) pipelines often fail to provide sufficient context, resulting in lower answer quality \cite{gao2023precise}. To improve support for highly novel or out-of-scope questions, we introduce an adaptive thresholding strategy. In this approach, if the initial retrieval does not yield any relevant content above the set similarity threshold, the system automatically relaxes the threshold in discrete steps (-0.1 at a time) until some content is retrieved or a minimum threshold is reached. This iterative process helps maximize the chances of finding relevant content for novel or previously unseen questions. 

We evaluate the adaptive thresholding strategy using the \textbf{Unseen Question Set}, which consists of questions that do not appear in the \textbf{RAG Knowledge Base} and are posted after the release date of the generation model. This setup is intended to simulate real-world scenarios where developers encounter entirely new or previously unseen questions that are not covered by the existing knowledge base.

\subsection{Applying the Approach across Different LLMs}
To assess the generalizability and practical impact of our optimal RAG pipeline, we apply it to multiple open-source LLMs to determine whether it consistently improves answer quality over standard zero-shot prompting. Specifically, we assess the pipeline using \textit{Granite-3.1-8B-Instruct}~\cite{mishra2024granite}, \textit{Mistral-7B-Instruct-v0.3}~\cite{mistral2024mistral7b}, \textit{Qwen3-8B}~\cite{yang2025qwen3}, and \textit{LLaMA-3.1-8B-Instruct}. All selected models have been extensively evaluated in recent studies and have demonstrated strong abilities in code generation, software comprehension, and developer-focused question answering tasks~\cite{kumar2025codecapbench,yang2025qwen3,mirza2025stratified}. For each model, we generate responses under two settings: (1) using our optimal RAG configuration and (2) using a baseline zero-shot prompting approach without any retrieved context. The \textbf{Synthetic Question Set} serves as the evaluation benchmark for this comparison.

We score every answer with the LLM-as-a-Judge model as outlined in Section~\ref{sec:evaluation}.
In addition, we conduct statistical significance testing to measure the differences between RAG-augmented and Stack Overflow accepted answers or zero-shot responses. We run the \textit{all-mpnet-base-v2} embedding, \textit{LLaMA-3.1-8B-Instruct}, \textit{Granite-3.1-8B-Instruct} and \textit{Mistral-7B-Instruct-v0.3}, and \textit{Qwen3-8B} models locally on an \textit{(NVIDIA)} A100 GPU with 80 GB of memory. 


\input{prompts/response_generation}

%% file: tables/rag_pipeline.tex
\begin{table*}[t]
\centering
\caption{Summary of RAG pipeline variants. All pipelines share the same
similarity-threshold range (0.1–0.9).}
\label{tab:rag_variants}
\begin{tabular}{llllcl}
\toprule
\textbf{ID} & \textbf{Pipeline} &
\textbf{Retrieval Target} &
\makecell{\textbf{Content}\\\textbf{Granularity}} &
\makecell{\textbf{Similarity}\\\textbf{Threshold}} &
\textbf{Notes} \\
\midrule
QB1 & Question-Based 1 & Direct   & Sentence     &
\multirow{7}{*}{0.1--0.9} &
Basic sentence-level retrieval \\
QB2 & Question-Based 2 & Direct   & Full answer  &  &
Basic full-answer retrieval \\
QB3 & Question-Based 3 & Indirect & Sentence     &  &
Indirect retrieval via similar questions \\
QB4 & Question-Based 4 & Indirect & Full answer  &  &
Indirect full-answer retrieval via similar questions \\
HB1 & HyDE-Based 1     & Direct   & Full answer  &  &
Basic HyDE \\
HB2 & HyDE-Based 2     & Direct   & Sentence     &  &
Basic HyDE, sentence level \\
HYB & QB + HyDE-Based  & Indirect & Full answer  &  &
Full hybrid pipeline \\
\bottomrule
\end{tabular}
\end{table*}

%% file: tables/similarity_threshold.tex
\begin{table}[ht!]
\centering
\caption{LLM-as-a-Judge mean scores by pipeline and similarity threshold.
Bold marks the best score in each row. An en-dash (–) indicates that the pipeline
retrieved no context at the specified threshold.}
\label{tab:threshold_scores_singlecol}
\small
\begin{tabular}{c|cccc|cc|c}
\toprule
\textbf{\makecell{Thres- \\ hold}} & 
\multicolumn{4}{c|}{\textbf{Question-Based (QB)}} & 
\multicolumn{2}{c|}{\makecell{\textbf{HyDE-Based} \\ \textbf{(HB)}}} & 
\textbf{Hybrid} \\ 
\cmidrule(lr){2-5} \cmidrule(lr){6-7}
& QB1 & QB2 & QB3 & QB4 & HB1 & HB2 & \\ 
\midrule
0.1 & 5.63 & 5.59 & 5.33 & 5.40 & \textbf{6.00} & 5.43 & 5.40 \\
0.2 & 5.55 & 4.85 & 5.26 & 5.19 & \textbf{5.92} & 5.41 & 5.21 \\
0.3 & 5.59 & 4.00 & 5.37 & 2.50 & \textbf{5.94} & 5.39 & 5.41 \\
0.4 & 5.51 & --   & 5.33 & --   & \textbf{5.82} & 5.28 & 5.48 \\
0.5 & 5.58 & --   & 5.36 & --   & \textbf{5.95} & 5.38 & 5.47 \\
0.6 & 5.66 & --   & 5.69 & -- & \textbf{5.94} & 5.49 & 5.43 \\
0.7 & 5.89 & --   & 5.86 & --   & \textbf{6.05} & 5.61 & 5.59 \\
0.8 & 6.01 & --   & 5.63 & --   & \textbf{6.39} & 5.80 & 5.25 \\
0.9 & --   & --   & --   & --   & \textbf{7.00} & 6.00 & -- \\
\midrule
\textit{Mean} & 5.68 & 4.81 & 5.48 & 4.36 & \textbf{6.11} & 5.53 & 5.41 \\
\bottomrule
\end{tabular}
\end{table}

%% file: prompts/response_generation.tex

\begin{figure}[t]
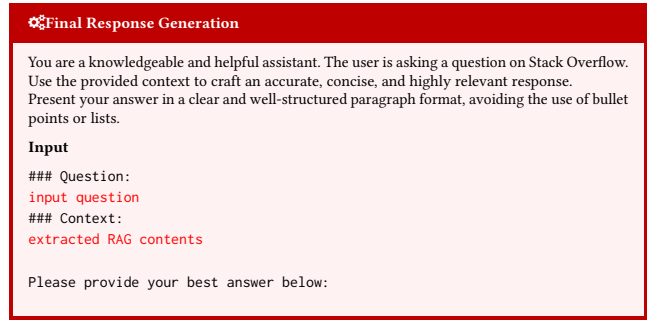

\begin{tcolorbox}[promptstyle,
  title={\faCogs Final Response Generation},colback=red!5!white,colframe=red!75!black,]

You are a knowledgeable and helpful assistant. The user is asking a question on Stack Overflow. \\
Use the provided context to craft an accurate, concise, and highly relevant response. \\
Present your answer in a clear and well-structured paragraph format, avoiding the use of bullet points or lists. 

\medskip
\textbf{Input}\par
\begin{lstlisting}[basicstyle=\ttfamily\scriptsize,frame=none]
### Question:
(*@\textcolor{red}{input question}@*)
### Context:
(*@\textcolor{red}{extracted RAG contents}@*)

Please provide your best answer below:
\end{lstlisting}
\end{tcolorbox}
\vspace{-1.2\baselineskip}
\caption{The prompt is used to generate the final response to input question by combining the extracted content from RAG pipeline.
Tags in \textcolor{red}{red} are placeholders.}
\label{fig:response_generation}
\end{figure}

%% file: sections/5_rq.tex
\section{Research Questions} 
\label{sec:rqs}

In this section, we provide the motivation, approach, and results of our research questions.\\
\noindent\textbf{\rqone}

\subsubsection*{\textbf{Motivation}}
Although RAG is widely adopted to improve the outputs of LLMs, its effectiveness is highly dependent on the design of its components \cite{su2024towards,li2025enhancing}. In this RQ, we investigate how different RAG design decisions affect the quality of the generated responses. Our goal is to determine the most effective RAG pipeline. In addition, subsequent experiments (e.g., RQ2) apply dynamic thresholding that progressively lowers the threshold to ensure every question receives context, it is important to evaluate how each pipeline performs across the full range of retrieval thresholds. This ensures that the selected configuration remains robust when coupled with dynamic thresholding in subsequent experiments.

\subsubsection*{\textbf{Approach}}
To evaluate the effectiveness of different RAG pipeline configurations, we assess 63 pipeline variants (summarized in Table~\ref{tab:rag_variants}) using the Synthetic Question Set, which includes 385 Stack Overflow questions whose corresponding answers are present in the RAG knowledge base (referred to as \textit{Seen Cases}). Each pipeline is evaluated along two key dimensions:

\textbf{(1) Answer quality.} We assess the quality of the generated responses based on three criteria: helpfulness (whether the response addresses the user query), correctness (technical accuracy), and level of detail (completeness and depth). Each response is scored from 1 to 10 using an LLM-as-a-Judge prompt (Figure~\ref{fig:rq1_llm_judge}). To examine how similarity threshold affects quality, we conduct a sensitivity analysis by varying the cosine similarity threshold from 0.1 to 0.9 across all RAG pipelines. This allows us to identify the most effective pipeline and optimal threshold setting. 


\textbf{(2) Retrieval Coverage.} We analyze how often each pipeline retrieves relevant content across the different similarity thresholds. This is computed using the same Synthetic Question Set, which is limited to the \textit{Seen Cases}. An effective pipeline should retrieve relevant content for as many Seen Cases as possible, as higher coverage reflects better stability and broader applicability. Retrieval Coverage is calculated as the proportion of questions for which relevant content is successfully extracted out of the total number of cases (385). However, coverage is not perfect—some questions still fail to retrieve relevant content even with the optimal pipeline and threshold. To address these cases, we explore the dynamic thresholding strategy in RQ2.

\begin{figure}[t]
    \centering
\includegraphics[width=0.5\textwidth]{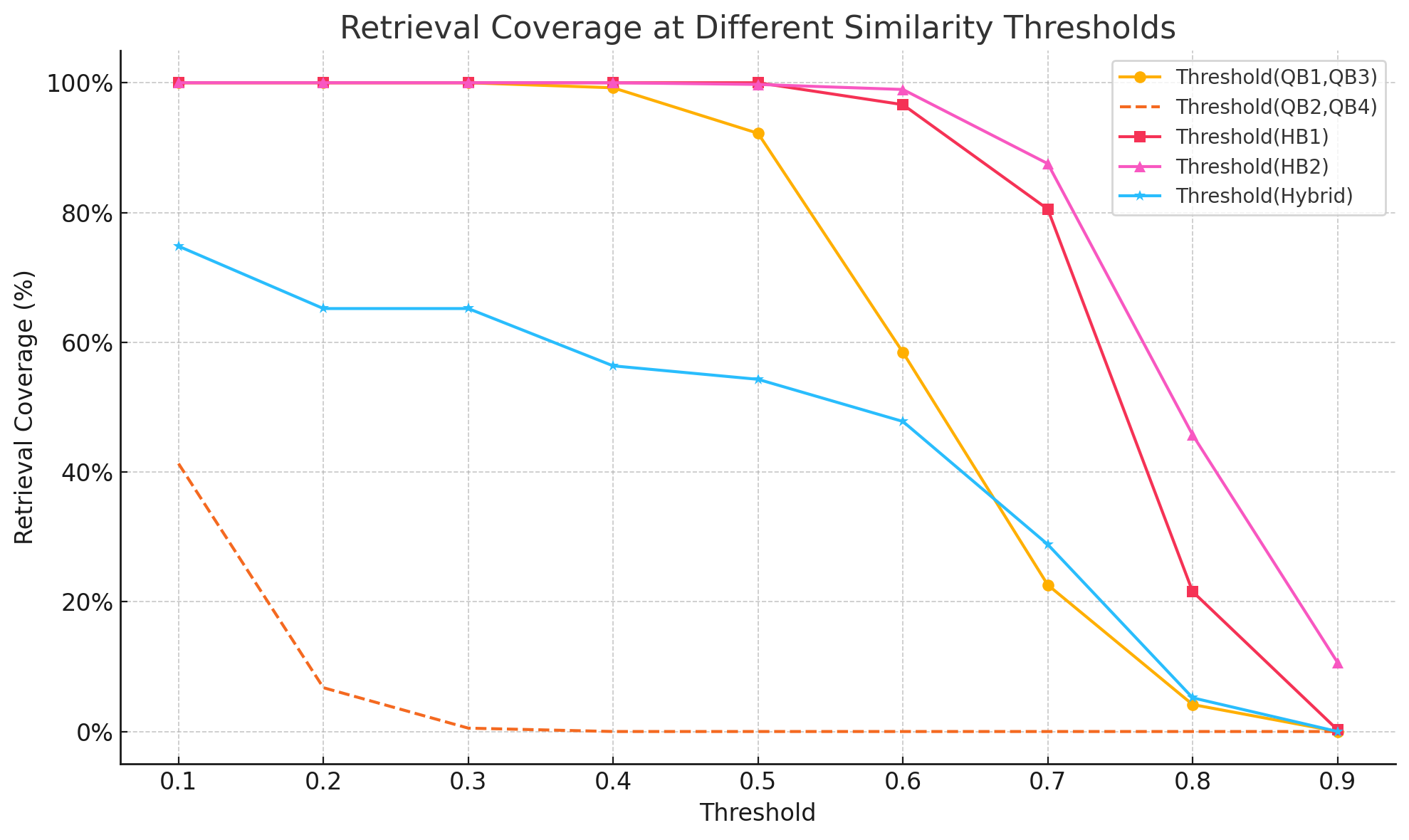}
    \caption{Percentage of retrieval coverage with retrievable context across varying similarity thresholds. Each line represents a different RAG pipeline. As the threshold increases, retrieval becomes more selective, reducing the number of questions with matching context.}
    \label{fig:sim_threshold}
\end{figure}

We compute LLM-as-a-Judge scores only for the Seen Cases where the pipeline successfully retrieves relevant content, excluding questions without retrieval from scoring in this RQ. Table~\ref{tab:threshold_scores_singlecol} summarizes the answer quality scores across all pipeline variants, while Figure~\ref{fig:sim_threshold} shows the percentage of Seen Cases with relevant content retrieved at each similarity threshold. Both answer quality and retrieval coverage are considered when selecting the optimal pipeline. 

\input{prompts/rq1_llm_judge.tex}

\subsubsection*{\textbf{Results:}}


\textbf{HB1, which uses direct retrieval over full answers with HyDE-generated queries, is the most effective pipeline, consistently outperforming others in both answer quality and retrieval coverage.} As shown in Table~\ref{tab:threshold_scores_singlecol}, HB1 achieves the highest average LLM-as-a-Judge score across all similarity thresholds (mean = 6.05), with a maximum score of 7 out of 10 at threshold 0.9. It also maintains strong retrieval coverage, retrieving relevant content for 80\% of Seen Cases at threshold 0.7. 
While HB2, which applies HyDE-based queries over sentence-level content, achieves similar coverage, its average quality score is lower (mean = 5.53), suggesting that answer-level granularity contributes to HB1’s better performance. In contrast, QB2, which performs direct retrieval over full answers using the original question as the query, shows weaker performance on both metrics, with significantly lower coverage and a mean score of only 4.81. These results confirm that HB1 offers the most robust balance between answer quality and retrieval coverage, making it the optimal configuration for downstream use in dynamic thresholding scenarios (RQ2).


\textbf{Higher thresholds generally improve generation quality but reduce the percentage of \textbf{\textit{Seen Cases}}.} For example, HB1 achieves the highest score (7.00) at threshold 0.9,  but retrieves relevant content for fewer than 1\% of Seen Cases, limiting its practical value. In contrast, HB1 at threshold 0.7 provides the best balance. It reaches a strong mean score of 6.05 while maintaining high \textit{retrieval coverage} of \textit{Seen Cases} percentage. This configuration offers the best trade-off between \textit{quality of answer} and \textit{retrieval coverage} among the pipelines. The results also reveal that even within Seen Cases, no pipeline achieves perfect coverage, motivating our subsequent experiment.

\medskip
\noindent\textbf{\rqtwo}
\subsubsection*{\textbf{Motivation}}
In practice, developers often ask questions that do not closely match any entry in a RAG knowledge base. Since RAG pipelines depend on retrieving relevant context \cite{gao2023precise}, the absence of such content limits the effectiveness, which leads the model to rely on training knowledge. In RQ1, even when evaluating on a synthetic question set sampled from the knowledge base, our optimal pipeline (HB1) retrieved relevant content for about 80\% of questions at threshold 0.7. This highlights the limitation that high-threshold retrieval cannot guarantee full coverage. In this RQ, we explore whether it is possible to improve coverage and answer quality on a more realistic testing set consisting of Stack Overflow questions posted after both the knowledge base cutoff and the release of the generation model.

\input{prompts/rq2_llm_judge.tex}

\subsubsection*{\textbf{Approach}} 
To address RQ2, we evaluate our method on the \textbf{Unseen Question Set}. Each question in the \textbf{Unseen Question Set} is paired with its original accepted answer from Stack Overflow, allowing us to compare the performance between the generated responses and the SO accepted answers.

Then, we apply the optimal pipeline identified in RQ1 (HB1), to retrieve context and generate an answer. The retrieval process starts with the highest similarity threshold. 
If no relevant content is retrieved, we apply the adaptive thresholding strategy.
Each answer is then evaluated using the LLM-as-a-Judge prompt (Figure~\ref{fig:rq2_llm_judge}), which generates a score based on helpfulness, correctness, and detail. These scores are used to compare the quality of generated responses against accepted Stack Overflow answers. To statistically compare the quality of our method against accepted answers, we use the Mann–Whitney U test ($p$-value) and report effect size using Cliff’s delta.


\subsubsection*{\textbf{Results}} 

Table~\ref{tab:dynamic_threshold_results} summarizes the performance of the adaptive thresholding strategy across similarity thresholds ranging from 0.9 to 0.5. 

\textbf{We observe a clear trade-off between quality and coverage.} At the highest threshold (0.9), the pipeline retrieves context for only 0.7\% of questions but achieves the highest response quality (mean = 6.44, median = 7.0), outperforming the corresponding accepted answers (mean = 5.03, median = 5.0). As the threshold is lowered, coverage increases substantially, reaching 31.9\% at 0.8 and 76.5\% at 0.7. Up to this threshold (0.7), generated responses consistently outperform accepted answers, with statistically significant differences and medium to small effect sizes (Cliff’s delta ranging from 0.41 to 0.28, $p <$ 0.001).

\textbf{While adaptive thresholding achieves full coverage by 0.5, it comes at the cost of reduced but reasonable  answer quality.} At thresholds 0.6 and 0.5, generated responses are outperformed by accepted answers, with Cliff’s delta values of $-0.29$ and $-0.44$, indicating small to medium negative effect sizes. It is important to note that the accepted answer scores at each threshold are not affected by the threshold itself; they simply reflect the scores for the subset of questions retrieved at that level. Despite the decline, the generated answers remain within the range of scores observed for accepted answers across the dataset (means between 4.69 and 6.34), with median scores equal to 6.0. This indicates that even when the retrieved context is only weakly related, the generated responses maintain a reasonable level of quality.

\textbf{When considering the entire Unseen Question Set, the adaptive thresholding strategy leads to a statistically significant overall improvement. }The weighted mean score for generated answers is 5.76 (median = 6.0), compared to 5.04 (median = 5.0) for accepted answers. The difference is statistically significant (Mann- Whitney U, $p < 10^{-82}$) with a small positive effect size (Cliff's delta = 0.21).
Overall, these results demonstrate that dynamic thresholding reliably improves or matches the quality of accepted answers, while guaranteeing full retrieval coverage for previously unseen questions.

\textbf{Qualitative insights.} Figure \ref{fig:rq2_qual_example} illustrates a typical unseen question outcome. The accepted answer receives an LLM-Judge score of 5/10 because it only lists the basic \texttt{getInt(int)} call and provides minimal justification.
On the other hand, the HB1 response scores 7/10 because it lists three alterbative APIs (\texttt{getInt}, \texttt{getObject}, \texttt{getLong/Short}) and offers additional guidance on when each is appropriate. This example illustrates the pattern observed in our quantitative results: HB1 provides more context and usage advice than the accepted answer on SO.

\input{figures/RQ2-qual-example}

\input{tables/dynamic_threshold.tex}

\medskip
\noindent\textbf{\rqthree}

\subsubsection*{\textbf{Motivation}}
While our proposed RAG pipeline has demonstrated effectiveness with one language model, it is unclear whether this performance (i) transfers to models with different architectures and pre-training, and (ii) still outperforms a strong zero-shot baseline for each model.
By evaluating HB1 on multiple LLMs and comparing its output with the models’ own zero-shot responses, we can measure both robustness (does the pipeline generalize?) and added value (does retrieval still help when the underlying model changes?).


\subsubsection*{\textbf{Approach}}
To evaluate the generalizability and practical impact of our optimal RAG pipeline, we apply it to multiple LLMs to test whether it improves answer quality over standard zero-shot prompting. We select several widely used models for this comparison, including \textit{Granite-3.1-8B-Instruct}, \textit{Mistral-7B-Instruct-v0.3}, \textit{Qwen3-8B} and \textit{LLaMA-3.1-8B-Instruct} used in previous RQs.

For each model, we generate answers to the 385 questions in the \textbf{Synthetic Question Set} using two configurations: (1) our optimized RAG pipeline (HB1) with dynamic thresholding from 0.9 to 0.1, and (2) a standard zero-shot prompting baseline without retrieval. All responses are assessed using the same LLM-as-a-Judge framework as in Figure\ref{fig:rq1_llm_judge}, with a focus on helpfulness, correctness, and level of detail. To statistically compare the effectiveness of each approach, we use the Wilcoxon signed-rank test \cite{wilcoxon1945individual}, which is well-suited for evaluating paired scores from HB1 and zero-shot responses for each LLM. This non-parametric test is appropriate because it directly compares paired samples without assuming the score distributions are normal. 

\vspace{-1em}
\subsubsection*{\textbf{Results}} 
\begin{figure}[h]
  \centering
\caption{Comparison of answer score distributions between zero-shot prompting and the HB1 RAG pipeline across multiple LLMs}
  \includegraphics[width=0.45\textwidth]{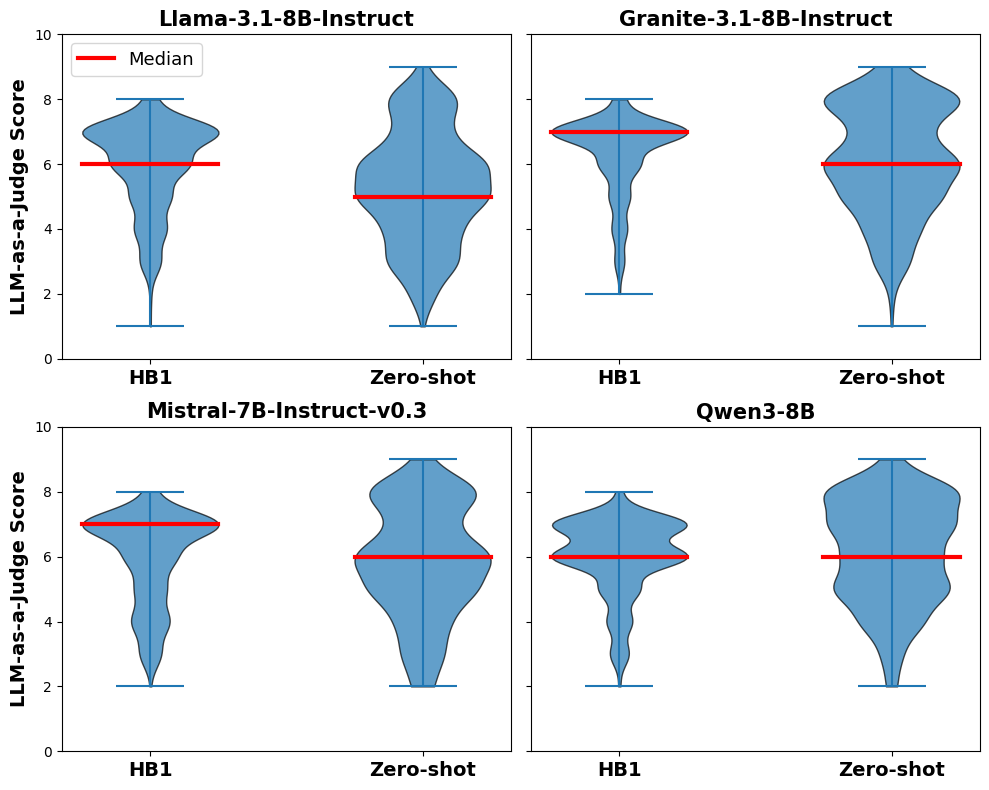}
  \label{fig:rq3_comparision}
\end{figure}

Figure~\ref{fig:rq3_comparision} presents the performance differences between zero-shot prompting and HB1 across all tested LLMs.
\textbf{Our optimal RAG pipeline (HB1) consistently improves or matches answer quality across three evaluated LLMs.} For LLaMA-3.1-8B-Instruct, HB1 achieves a mean score of 5.95 (median 6.0), surpassing the zero-shot baseline (mean 5.31, median 5.0) with statistical significance (\textit{Wilcoxon p-value: $1 \times 10^{-2}$}). Granite-3.1-8B-Instruct shows a similar pattern: HB1 reaches a mean of 6.31 (median 7.0), compared to 6.14 (median 6.0) for zero-shot (\textit{p-value: $3 \times 10^{-3}$}). For Mistral-7B-Instruct-v0.3, HB1’s mean is 5.96 (median 7.0), again outperforming zero-shot (mean 5.83, median 6.0) with a highly significant difference (\textit{p-value: $8.8 \times 10^{-6}$}).

\textbf{For Qwen3-8B, HB1 does not improve over zero-shot prompting, showing that better-trained models benefit less from retrieval augmentation.} In contrast, for Qwen3-8B, zero-shot prompting yields a slightly higher mean score (6.15) than HB1 (5.97), though both achieve a median of 6.0 and no statistically significant difference is observed. 


\textbf{While mean scores offer a high-level comparison, the distribution reveals how HB1 improves not just typical answer quality but also output consistency}, as seen in Figure~\ref{fig:rq3_comparision}.
First, for models such as Mistral, LLaMA, and Granite, HB1 reduces the frequency of low-quality outputs (scores $<3$) and narrows the overall distribution. This indicates not only higher average scores, but also increased answer consistency. For Mistral and LLaMA, we note an observable upward shift in the distribution. 
Second, in the case of Qwen3-8B, both HB1 and zero-shot exhibit somewhat similar wide distributions centered around a median of 6. This suggests that stronger models with broader pretraining may benefit less from retrieval, as they already encode much of the required knowledge.

Overall, these results confirm that our optimal RAG pipeline - HB1 not only generalizes well across open source LLMs but also enhances answer quality and reliability in most cases, supporting its practical value for technical Q\&A.

%% file: prompts/rq1_llm_judge.tex
\begin{figure}[t]
\begin{tcolorbox}[promptstyle,
  title={\faBalanceScale\; Seen data Judge Prompt}]

\textbf{System}. We would like your feedback on several approaches
to the question in \texttt{[QUESTION TITLE]}.  
Rate the helpfulness, accuracy, and level of detail of each response.  
If a response contains an incomplete code snippet, deduct from the score.  
Give each approach a score from 1–10; higher is better.

\medskip
\textbf{Output format}. One line with the scores, separated by commas.

\medskip
\textbf{Input}\par
\begin{lstlisting}[basicstyle=\ttfamily\scriptsize,frame=none]
[QUESTION TITLE]
(*@\textcolor{red}{question title}@*)
[THE START OF APPROACH 1 ANSWERS]
(*@\textcolor{red}{response 1}@*)
[THE END OF APPROACH 1 ANSWERS]
[THE START OF APPROACH 2 ANSWERS]
(*@\textcolor{red}{response 2}@*)
[THE END OF APPROACH 2 ANSWERS]
               ...
[THE START OF APPROACH n ANSWERS]
(*@\textcolor{red}{response n}@*)
[THE END OF APPROACH n ANSWERS]             
\end{lstlisting}
\end{tcolorbox}
\vspace{-1.2\baselineskip}
\caption{Prompt used by the LLM-as-Judge to evaluate generated answers for helpfulness, accuracy, and detail across all RAG pipelines at a single threshold.
Tags in \textcolor{red}{red} are placeholders replaced at evaluation time.}
\label{fig:rq1_llm_judge}
\end{figure}

%% file: prompts/rq2_llm_judge.tex
\begin{figure}[t!]
\begin{tcolorbox}[promptstyle,
  title={\faBalanceScale\; Unseen data Judge Prompt}]

\textbf{System}. We would like your feedback on several approaches
to the question in \texttt{[QUESTION TITLE]}.  
Rate the helpfulness, accuracy, and level of detail of each response.  
If a response contains an incomplete code snippet, deduct from the score.  
Give each approach a score from 1–10; higher is better.

\medskip
\textbf{Output format}. One line with the scores, separated by commas.

\medskip
\textbf{Input}\par
\begin{lstlisting}[basicstyle=\ttfamily\scriptsize,frame=none]
[QUESTION TITLE]
(*@\textcolor{red}{question title}@*)
[THE START OF APPROACH 1 ANSWERS]
(*@\textcolor{red}{response 1}@*)
[THE END OF APPROACH 1 ANSWERS]
[THE START OF APPROACH 2 ANSWERS]
(*@\textcolor{red}{response 2}@*)
[THE END OF APPROACH 2 ANSWERS]
\end{lstlisting}
\end{tcolorbox}
\vspace{-1.2\baselineskip}
\caption{Prompt used by the LLM-as-Judge to score generated answers
and the accepted answer on helpfulness, accuracy, and detail.
Tags in \textcolor{red}{red} are placeholders replaced at evaluation time.}
\label{fig:rq2_llm_judge}
\end{figure}

%% file: figures/RQ2-qual-example.tex
\begin{figure}[t]
\centering
\begin{minipage}[t]{0.36\linewidth}
\small
\textbf{Query}\\
\emph{What methods can be used to transform a \texttt{ResultSet} into an actionable integer value?}

\vspace{4pt}
\textbf{Accepted answer (LLM-Judge = 5)}\par
\begin{lstlisting}
A ResultSet has a series of getXYZ(int) methods to retrieve
columns from it by their relative index and corresponding
getXYZ(String) methods to retrieve those columns by their alias.

In your case, using the index variant  getInt(int) would be
the easiest:
recordNumber = resultSet.getInt(1);
\end{lstlisting}
\end{minipage}\hfill
\begin{minipage}[t]{0.56\linewidth}
\small
\textbf{Adaptive HB1 (LLM-Judge = 7)}\par
\begin{lstlisting}
To transform a ResultSet into an actionable integer value,  
you can use one of the following methods:

- getInt(int column): [Basic explanation omitted.]
- getObject(int column) followed by casting to Integer:  
  [Details about usage and casting omitted.]
- getLong(int column) or getShort(int column):  
  [Explanation of method differences omitted.]

Here's an example of how to use these methods:  
[Sample code omitted.]

Note that getInt(int column) is generally the recommended  
approach when working with ResultSets in Java.  
[Further explanation omitted.]
\end{lstlisting}
\end{minipage}
\caption{Qualitative example from the \emph{Unseen Question Set}.  
HB1 scores 7/10, outperforming the accepted answer (5/10).}
\label{fig:rq2_qual_example}
\end{figure}

%% file: tables/dynamic_threshold.tex
\begin{table*}[h]

\caption{Adaptive–threshold performance on the \textit{Unseen Question Set} (5 510 questions).  
For each threshold we report answer quality (generated vs. accepted), significance tests, and retrieval coverage.  
Cumulative coverage reaches 100\% once the threshold reaches 0.5.}
\label{tab:dynamic_threshold_results}
\setlength{\tabcolsep}{4pt}
\begin{tabular}{ccccccrrrrr}
\toprule
\multirow{2}{*}{\textbf{Threshold}}
      & \multirow{2}{*}{\textbf{\# Questions}} 
      & \multicolumn{2}{c}{\textbf{Mean}} 
      & \multicolumn{2}{c}{\textbf{Median}}
      & \multicolumn{4}{c}{\textbf{Statistical test}}
      & \multirow{2}{*}{\textbf{Coverage (\%)}} \\ 
\cmidrule(lr){3-4}\cmidrule(lr){5-6}\cmidrule(lr){7-10}
      &           & Generated & Accepted & Generated & Accepted 
      & \textbf{U} & \textbf{p} & \boldmath$\Delta$ & \textbf{ES} & \\ \midrule
0.9 & 39      & \textbf{6.44} & 5.03 & \textbf{7.0} & 5.0 & 1 071.5 & 0.0015 &  0.41 & med. &  \phantom{0}0.7 (+0.7) \\
0.8 & 1 720   & \textbf{6.20} & 4.91 & \textbf{7.0} & 5.0 & 2 037 269 & <0.001 &  0.38 & med. &  31.9 (+31.2) \\
0.7 & 2 456   & \textbf{5.60} & 4.69 & \textbf{6.0} & 4.0 & 3 874 331 & <0.001 &  0.28 & small &  76.5 (+44.6) \\
0.6 & 1 181   & 5.50 & \textbf{6.10} & 6.0 & \textbf{7.0} &   494 031 & <0.001 & –0.29 & small &  97.9 (+21.4) \\
0.5 & 108     & 5.13 & \textbf{6.34} & 6.0 & \textbf{7.0} &     3 260 & <0.001 & –0.44 & med.  & 100.0 (+2.1) \\ \bottomrule
\multicolumn{11}{l}{\small U = Mann–Whitney statistic; 
$p$ = two-sided p-value; $\Delta$ = Cliff’s delta; ES = effect-size label.}
\end{tabular}

\end{table*}

%% file: sections/6_discussion.tex
\section{Discussion}
\label{sec:implications}
\subsection{Qualitative Insights}

Results from the research questions show that our optimal RAG pipeline consistently improves or matches the accepted answers on Stack Overflow dataset and the zero-shot prompting across all evaluated LLMs.   
However, numeric gains alone do not fully capture the practical value of the proposed  RAG pipeline. 
To better understand the nature of these improvements, we conduct a focused qualitative review of the model outputs.

We randomly sample 20 questions (5 per model) from the KB-Synthetic Q-Set where HB1. For each question, we compare the answers and highlight common improvements, such as increased detail, clearer structure, or usefulness. We then consolidate our observations to identify common qualitative themes that illustrate how HB1 enhances the response value beyond the observed few point margin in the LLM-As-A-Judge score.

\smallskip
\noindent\textbf{Cases where HB1 improves response quality:} Among the examples analyzed, HB1 outperformed zero-shot prompting in 75\% of the cases. Score differences range from +2 to +6, with a mean of 2.3 and a median of 3.0. The qualitative review reveals consistent improvements in the utility and presentation of the content.

Specifically, HB1 answers more frequently includes:
    \begin{itemize}[leftmargin=5.5pt,itemindent=0pt,labelsep=0.5em] 
        \item \textbf{Best-practice use of APIs.} HB1 answers more frequently use concise, recommended solutions that align with best practices in a given programming ecosystem. For example, in a question about assigning group indices in a Pandas DataFrame, HB1 use the best-practice one-liner solution \texttt{df['id'] = df.groupby('col')} \texttt{.cumcount() + 1}, while the zero-shot answer suggests manually looping through rows, an approach that is less efficient and harder to maintain.
        \item \textbf{Richer contextualization.} Compared to zero-shot answers, HB1 more often includes brief explanations of \emph{why} a solution works or when it should be used. For instance, in a question about \texttt{plt.axis(`equal')} in Matplotlib, HB1 explains that this ensures equal scaling on both axes (useful for plotting circles or squares accurately) and shows how to adjust the figure layout to avoid distortion, which are details missing in the zero-shot answer.
        \item \textbf{Edge-case handling.} HB1 is also more likely to mention failure scenarios or constraints. For example, when answering a question about limiting checkbox selection to three options using jQuery, the HB1 response includes logic to disable unchecked boxes once the user reaches the limit. This type of input validation is rarely included in zero-shot answers but is critical for correct user interaction in production environments.
    \end{itemize}
Table~\ref{tab:qual_win} lists five representative examples where HB1 adds value compared to the baseline.
\begin{table*}[t]
\centering\small
\caption{Examples where HB1 significantly outperforms zero-shot.}
\label{tab:qual_win}
\setlength{\tabcolsep}{4pt}
\begin{tabular}{p{3.1cm} p{1.1cm} p{3.3cm} p{3.3cm} p{4.0cm}}
\toprule
\textbf{Question (shortened)} & \textbf{Model} &
\textbf{Zero-shot focus} &
\textbf{HB1 focus} &
\textbf{Why HB1 wins} \\ \midrule

Counting series in Pandas          & Mistral &
Uses a manual row loop to assign group indices. &
Applies the concise \texttt{groupby().cumcount() + 1} idiom. &
HB1 provides the correct and efficient one-liner standard in real-world Pandas use. \\[4pt]

jQuery execution order             & LLaMA   &
Mentions only \texttt{.each()} as an option. &
Lists five sequencing patterns (e.g., \texttt{\$().ready()}, promises, \texttt{setTimeout}). &
HB1 offers broader and more practical coverage for managing execution flow. \\[4pt]

Zero-padding integers in Java      & LLaMA   &
Shows \texttt{String.format()} without explanation. &
Includes both \texttt{String.format()} and manual padding with width logic. &
HB1 is more complete and explains when to use each approach. \\[4pt]

\texttt{plt.axis('equal')} \\behaviour & Qwen    &
Calls \texttt{plt.axis('equal')} but omits layout implications. &
Adds \texttt{set\_aspect('equal', adjustable='box')} and explains distortion issues. &
HB1 provides a more robust solution with better layout control. \\[4pt]

Vue \texttt{v-model} with props    & Mistral &
Uses basic prop binding but omits two-way support. &
Implements the full computed getter/setter pattern with \texttt{@input}. &
HB1 shows the correct two-way binding pattern for props. \\ \bottomrule
\end{tabular}
\end{table*}

\noindent\textbf{Cases where zero-shot performs better:} In 25\% of the cases from the random sample, zero-shot prompting outperformed HB1. These cases are more likely questions that require conceptual clarity rather than code synthesis. We observe that retrieval either introduces off-topic content or leads the model to address a broader variant of the task, missing the specific user intent. Table~\ref{tab:qual_loss} shows examples of the questions and the observed issues. In two AngularJS questions, HB1 is misled by retrieved content and generates off-topic answers about service patterns. In the third case, HB1 overgeneralizes a question about removing duplicate permutations, solving a more complex variant than intended. In the three examples, zero-shot responses were better aligned with the user's intent.

\begin{table*}[t]
\caption{Examples where zero-shot significantly outperforms HB1.}
\label{tab:qual_loss}
\small
\setlength{\tabcolsep}{4pt}
\begin{tabular}{p{2.7cm}p{1.2cm}p{3.5cm}p{3.5cm}p{4.2cm}}
\toprule
\textbf{Question (shortened)} & \textbf{Model} &
\textbf{Zero-shot focus} &
\textbf{HB1 focus} &
\textbf{Why Zero-shot wins} \\ \midrule
Remove duplicate permutations of digit arrays &
Mistral &
Set-based \emph{one-liner} that correctly deduplicates permutations. &
Stringify() + \texttt{Set} technique suitable for flat arrays but not for \emph{arrays-of-arrays}. Problem is mis-interpreted. &
Zero-shot matches the stated task. HB1 over-generalizes and solves the wrong problem. \\ \addlinespace[2pt]

Curly-brace syntax in AngularJS &
Mistral &
Accurate explanation of \texttt{\{\{ \}\}} interpolation, security considerations, and \texttt{ngBind}. &
Discusses \texttt{factory} vs.\ \texttt{service} patterns resulting in a topic drift. &
Retriever fetches a semantically related but topically different answer which leads HB1 off-topic. \\ \addlinespace[2pt]

Curly-brace syntax in AngularJS  &
LLaMA &
Same concise and on-point tutorial as Mistral &
Repeats factory/service discussion. Still off-topic. &
Same retrieval mismatch. Zero-shot is aligned with the actual question. \\ \bottomrule
\end{tabular}
\end{table*}

To conclude, this analysis shows that HB1 outperforms zero-shot prompting in 75\% of sampled cases, offering code that is more aligned with best practices, contextual explanations, and edge-case handling. In the remaining 25\%, zero-shot responses are stronger, typically for concept-focused questions where retrieval led HB1 to go off-topic or to overgeneralize. 

\subsection{Implications}

\paragraph{Implications for practitioners.}
Our results demonstrate that combining HyDE-based retrieval with full-answer granularity and dynamic thresholding improves both the retrieval coverage and quality of generated answers. The HB1 pipeline improves the LLM-as-a-Judge scores while maintaining 100\% retrieval coverage on unseen questions. These improvements are most common in implementation-oriented questions, where HB1 produces code more in line with best practices, contextual detail, and input validation logic. Two configuration choices were particularly effective: (1) initializing retrieval at a high similarity threshold (e.g., 0.7) and lowering it only when no candidates are found, and (2) retrieving complete answers rather than individual sentences, as full answers more often include supporting rationale and edge-case handling. For concept-focused questions, retrieval was slightly less effective. As such, RAG based systems may benefit from a lightweight classifier to decide when retrieval should be skipped in favor of zero-shot prompting.

\paragraph{Implications for researchers.}
The study raises two research directions. First, we observe model-dependent variation when measuring the benefit of retrieval: Qwen-3-8B showed minimal gains from HB1, likely due to broader pretraining, while other models show important improvement. Future work should investigate when retrieval is beneficial based on model characteristics or question type. Second, most of the improvements observed in HB1, such as the best-practice use of APIs and secure coding practices, are unlikely to be captured by standard metrics commonly used in prior work (e.g., BLEU, ROUGE), which rely on lexical overlap. Our use of LLM-as-a-Judge enables an evaluation of aspects like completeness and practical utility. This highlights the need for evaluation methods that better reflect developer-relevant quality criteria, and motivates the development of metrics that assess answer usefulness beyond surface-level similarity.





%% file: sections/7_threats.tex
\section{Threats to validity}
\label{sec:threats_to_validity}
In this section, we discuss the threats to the validity of our study.

\textbf{Threats to construct validity} 
Our evaluation relies on LLM-as-a-Judge, which used and evaluated in recent studies, though it may still introduce bias, inconsistency, or hallucination. To mitigate these risks, we anonymize responses and conduct manual evaluations to assess the alignment between LLM-as-a-Judge scores and human judgments (see Section~\ref{sec:evaluation}).

\textbf{Threats to internal validity}
Data leakage is a significant concern in LLM-based evaluation, as test questions may overlap with the training data or retrieval corpus. To mitigate this risk, we generate synthetic questions for "seen" cases and construct an unseen test set using posts published after the LLM training cutoff.

\textbf{Threats to external validity} 
We focus on questions tagged with Java and Python from Stack Overflow, which may limit generalization to other domains, languages, or Q\&A platforms. However, as Java and Python are among the most popular and widely used programming languages, our findings are likely relevant to a broad range of developer scenarios.


%% file: sections/8_related_work.tex
\section{Related Work}
\label{sec:related_work}
Stack Overflow remains a central resource for developer knowledge, and its content has been widely leveraged to build datasets and systems for tasks such as answer retrieval, code snippet recommendation, and question summarization~\cite{zhang2022sosum,kou2022answer}. Many studies have mined accepted answers or high-quality responses from Stack Overflow to serve as gold standards for training and evaluating machine learning models~\cite{parvez2021retrieval,gao2023code}. In recent years, Retrieval-Augmented Generation (RAG) has emerged as a way to enhance large language models (LLMs) with domain knowledge—including API usage, troubleshooting, and community explanations—to improve developer support~\cite{mukherjee2025sosecure}. While vanilla RAG setups rely on the input query for retrieval~\cite{lewis2020retrieval}, this approach can struggle with vague or novel questions. To address these challenges, recent works have proposed enhancements such as Hypothetical Document Embedding (HyDE), which generates a pseudo-answer to improve retrieval alignment~\cite{wang2024coderag}, and filtering mechanisms to remove irrelevant content before generation~\cite{mukherjee2025sosecure}. Recent works such as RAGFix~\cite{mansur2024ragfix} and StackRAG~\cite{abrahamyan2024stackrag} have demonstrated the potential of combining Stack Overflow knowledge with RAG pipelines for tasks like code repair and developer Q\&A. However, the most existing approaches focus on direct query-based retrieval, which often fails on vague or novel queries and do not explicitly address pipeline generalization. In contrast, our work systematically evaluates a range of retrieval strategies—including hypothetical document embedding (HyDE)—as well as content granularity and adaptive thresholding, using large-scale Stack Overflow data. Through this analysis, we identify pipeline configurations that effectively handle novel questions and generalize across different open-source LLMs.
\label{related_work}

%% file: sections/9_conclusion.tex
\section{Conclusion}\label{sec:conclusion}
In this paper, we investigate adopting RAG to improve the LLMs’ capability to answer developers’ questions by constructing RAG using Stack Overflow posts with accepted answers as a knowledge base. By constructing a large-scale dataset of Java and Python Q\&A pairs and evaluating seven different RAG pipeline designs, we identify HyDE-Based 1 (HB1) as the most effective. HB1 generates a hypothetical answer for each question and retrieves full accepted answers based on this pseudo-answer for context, achieving a strong balance between answer quality and retrieval coverage across a range of similarity thresholds. To address questions with no close matches in the knowledge base, we introduce adaptive thresholding, which dynamically lowers the retrieval threshold to improve coverage for novel questions. To simulate real-world scenarios, we evaluate our approach on 5,510 Stack Overflow questions sampled from posts after the LLM's training cutoff, ensuring they were unseen during retrieval. At high similarity thresholds, our approach often exceeds the quality of the original accepted answers. 

We further test our pipeline across multiple LLMs, finding that our adaptive RAG approach robustly enhances answer quality or matches strong zero-shot baselines for most models. Overall, our findings demonstrate that the proposed optimal RAG pipeline, combined with adaptive thresholding, provides a practical and effective solution for delivering reliable, high-quality developer assistance on both familiar and novel questions. In future work, we plan to evaluate our approach on additional datasets and with proprietary models such as the GPT family from OpenAI.


%% file: main.bbl

\begin{thebibliography}{41}


\ifx \showCODEN    \undefined \def \showCODEN     #1{\unskip}     \fi
\ifx \showISBNx    \undefined \def \showISBNx     #1{\unskip}     \fi
\ifx \showISBNxiii \undefined \def \showISBNxiii  #1{\unskip}     \fi
\ifx \showISSN     \undefined \def \showISSN      #1{\unskip}     \fi
\ifx \showLCCN     \undefined \def \showLCCN      #1{\unskip}     \fi
\ifx \shownote     \undefined \def \shownote      #1{#1}          \fi
\ifx \showarticletitle \undefined \def \showarticletitle #1{#1}   \fi
\ifx \showURL      \undefined \def \showURL       {\relax}        \fi
\providecommand\bibfield[2]{#2}
\providecommand\bibinfo[2]{#2}
\providecommand\natexlab[1]{#1}
\providecommand\showeprint[2][]{arXiv:#2}

\bibitem[Abrahamyan and Fard(2024)]%
        {abrahamyan2024stackrag}
\bibfield{author}{\bibinfo{person}{Davit Abrahamyan} {and} \bibinfo{person}{Foutse~Khomh Fard}.} \bibinfo{year}{2024}\natexlab{}.
\newblock \showarticletitle{StackRAG Agent: Improving Developer Answers with Retrieval-Augmented Generation}. In \bibinfo{booktitle}{\emph{IEEE International Conference on Software Maintenance and Evolution (ICSME)}}.
\newblock
\urldef\tempurl%
\url{https://ieeexplore.ieee.org/document/10795043}
\showURL{%
\tempurl}


\bibitem[AI(2024a)]%
        {meta2024llama3}
\bibfield{author}{\bibinfo{person}{Meta AI}.} \bibinfo{year}{2024}\natexlab{a}.
\newblock \bibinfo{title}{LLaMA 3.1–8B-Instruct}.
\newblock \bibinfo{howpublished}{\url{https://huggingface.co/meta-llama/Meta-Llama-3-8B-Instruct}}.
\newblock


\bibitem[AI(2024b)]%
        {mistral2024mistral7b}
\bibfield{author}{\bibinfo{person}{Mistral AI}.} \bibinfo{year}{2024}\natexlab{b}.
\newblock \bibinfo{title}{Mistral-7B-Instruct-v0.3}.
\newblock \bibinfo{howpublished}{\url{https://huggingface.co/mistralai/Mistral-7B-Instruct-v0.3}}.
\newblock


\bibitem[Authors(2025)]%
        {icse2026_stackoverflow_rag}
\bibfield{author}{\bibinfo{person}{Anonymous Authors}.} \bibinfo{year}{2025}\natexlab{}.
\newblock \bibinfo{title}{ICSE-C2-Stack-overflow: Retrieval-Augmented Generation for Developer Questions}.
\newblock \bibinfo{howpublished}{\url{https://anonymous.4open.science/r/ICSE-C2-Stack-overflow-16E8/README.md}}.
\newblock
\newblock
\shownote{Accessed: July 2025}.


\bibitem[Bender et~al\mbox{.}(2021)]%
        {bender2021dangers}
\bibfield{author}{\bibinfo{person}{Emily~M Bender}, \bibinfo{person}{Timnit Gebru}, \bibinfo{person}{Angelina McMillan-Major}, {and} \bibinfo{person}{Shmargaret Shmitchell}.} \bibinfo{year}{2021}\natexlab{}.
\newblock \showarticletitle{On the dangers of stochastic parrots: Can language models be too big?}. In \bibinfo{booktitle}{\emph{Proceedings of the 2021 ACM conference on fairness, accountability, and transparency}}. \bibinfo{pages}{610--623}.
\newblock


\bibitem[Bommasani et~al\mbox{.}(2021)]%
        {bommasani2021opportunities}
\bibfield{author}{\bibinfo{person}{Rishi Bommasani}, \bibinfo{person}{Drew~A Hudson}, \bibinfo{person}{Ehsan Adeli}, \bibinfo{person}{Russ Altman}, \bibinfo{person}{Simran Arora}, \bibinfo{person}{Sydney von Arx}, \bibinfo{person}{Michael~S Bernstein}, \bibinfo{person}{Jeannette Bohg}, \bibinfo{person}{Antoine Bosselut}, \bibinfo{person}{Emma Brunskill}, {et~al\mbox{.}}} \bibinfo{year}{2021}\natexlab{}.
\newblock \showarticletitle{On the opportunities and risks of foundation models}.
\newblock \bibinfo{journal}{\emph{arXiv preprint arXiv:2108.07258}} (\bibinfo{year}{2021}).
\newblock


\bibitem[Cao et~al\mbox{.}(2021)]%
        {cao2021knowledgeable}
\bibfield{author}{\bibinfo{person}{Boxi Cao}, \bibinfo{person}{Hongyu Lin}, \bibinfo{person}{Xianpei Han}, \bibinfo{person}{Le Sun}, \bibinfo{person}{Lingyong Yan}, \bibinfo{person}{Meng Liao}, \bibinfo{person}{Tong Xue}, {and} \bibinfo{person}{Jin Xu}.} \bibinfo{year}{2021}\natexlab{}.
\newblock \showarticletitle{Knowledgeable or educated guess? revisiting language models as knowledge bases}.
\newblock \bibinfo{journal}{\emph{arXiv preprint arXiv:2106.09231}} (\bibinfo{year}{2021}).
\newblock


\bibitem[Cheng et~al\mbox{.}(2025)]%
        {cheng2025survey}
\bibfield{author}{\bibinfo{person}{Mingyue Cheng}, \bibinfo{person}{Yucong Luo}, \bibinfo{person}{Jie Ouyang}, \bibinfo{person}{Qi Liu}, \bibinfo{person}{Huijie Liu}, \bibinfo{person}{Li Li}, \bibinfo{person}{Shuo Yu}, \bibinfo{person}{Bohou Zhang}, \bibinfo{person}{Jiawei Cao}, \bibinfo{person}{Jie Ma}, {et~al\mbox{.}}} \bibinfo{year}{2025}\natexlab{}.
\newblock \showarticletitle{A survey on knowledge-oriented retrieval-augmented generation}.
\newblock \bibinfo{journal}{\emph{arXiv preprint arXiv:2503.10677}} (\bibinfo{year}{2025}).
\newblock


\bibitem[Cutting and Stephen(2021)]%
        {cutting2021comparative}
\bibfield{author}{\bibinfo{person}{Vineesh Cutting} {and} \bibinfo{person}{Nehemiah Stephen}.} \bibinfo{year}{2021}\natexlab{}.
\newblock \showarticletitle{Comparative review of java and python}.
\newblock \bibinfo{journal}{\emph{International Journal of Research and Development in Applied Science and Engineering (IJRDASE)}} \bibinfo{volume}{21}, \bibinfo{number}{1} (\bibinfo{year}{2021}).
\newblock


\bibitem[Gamage et~al\mbox{.}(2022)]%
        {gamage2022deepfakes}
\bibfield{author}{\bibinfo{person}{Dilrukshi Gamage}, \bibinfo{person}{Piyush Ghasiya}, \bibinfo{person}{Vamshi Bonagiri}, \bibinfo{person}{Mark~E Whiting}, {and} \bibinfo{person}{Kazutoshi Sasahara}.} \bibinfo{year}{2022}\natexlab{}.
\newblock \showarticletitle{Are deepfakes concerning? analyzing conversations of deepfakes on reddit and exploring societal implications}. In \bibinfo{booktitle}{\emph{Proceedings of the 2022 CHI conference on human factors in computing systems}}. \bibinfo{pages}{1--19}.
\newblock


\bibitem[Gao et~al\mbox{.}(2023a)]%
        {gao2023precise}
\bibfield{author}{\bibinfo{person}{Luyu Gao}, \bibinfo{person}{Xueguang Ma}, \bibinfo{person}{Jimmy Lin}, {and} \bibinfo{person}{Jamie Callan}.} \bibinfo{year}{2023}\natexlab{a}.
\newblock \showarticletitle{Precise zero-shot dense retrieval without relevance labels}. In \bibinfo{booktitle}{\emph{Proceedings of the 61st Annual Meeting of the Association for Computational Linguistics (Volume 1: Long Papers)}}. \bibinfo{pages}{1762--1777}.
\newblock


\bibitem[Gao et~al\mbox{.}(2023b)]%
        {gao2023code}
\bibfield{author}{\bibinfo{person}{Ziyu Gao}, \bibinfo{person}{Xin Xia}, \bibinfo{person}{David Lo}, \bibinfo{person}{John Grundy}, {and} \bibinfo{person}{Tian Zhang}.} \bibinfo{year}{2023}\natexlab{b}.
\newblock \showarticletitle{I Know What You Are Searching For: Code Snippet Recommendation from Stack Overflow Posts}.
\newblock \bibinfo{journal}{\emph{ACM Transactions on Software Engineering and Methodology (TOSEM)}} (\bibinfo{year}{2023}).
\newblock
\urldef\tempurl%
\url{https://dl.acm.org/doi/10.1145/3550150}
\showURL{%
\tempurl}


\bibitem[Gehman et~al\mbox{.}(2020)]%
        {gehman2020realtoxicityprompts}
\bibfield{author}{\bibinfo{person}{Samuel Gehman}, \bibinfo{person}{Suchin Gururangan}, \bibinfo{person}{Maarten Sap}, \bibinfo{person}{Yejin Choi}, {and} \bibinfo{person}{Noah~A Smith}.} \bibinfo{year}{2020}\natexlab{}.
\newblock \showarticletitle{Realtoxicityprompts: Evaluating neural toxic degeneration in language models}.
\newblock \bibinfo{journal}{\emph{arXiv preprint arXiv:2009.11462}} (\bibinfo{year}{2020}).
\newblock


\bibitem[Gilson et~al\mbox{.}(2024)]%
        {gilson2024enhancing}
\bibfield{author}{\bibinfo{person}{Aidan Gilson}, \bibinfo{person}{Xuguang Ai}, \bibinfo{person}{Thilaka Arunachalam}, \bibinfo{person}{Ziyou Chen}, \bibinfo{person}{Ki~Xiong Cheong}, \bibinfo{person}{Amisha Dave}, \bibinfo{person}{Cameron Duic}, \bibinfo{person}{Mercy Kibe}, \bibinfo{person}{Annette Kaminaka}, \bibinfo{person}{Minali Prasad}, {et~al\mbox{.}}} \bibinfo{year}{2024}\natexlab{}.
\newblock \showarticletitle{Enhancing Large Language Models with Domain-specific Retrieval Augment Generation: A Case Study on Long-form Consumer Health Question Answering in Ophthalmology}.
\newblock \bibinfo{journal}{\emph{arXiv preprint arXiv:2409.13902}} (\bibinfo{year}{2024}).
\newblock


\bibitem[Gomaa et~al\mbox{.}(2013)]%
        {gomaa2013survey}
\bibfield{author}{\bibinfo{person}{Wael~H Gomaa}, \bibinfo{person}{Aly~A Fahmy}, {et~al\mbox{.}}} \bibinfo{year}{2013}\natexlab{}.
\newblock \showarticletitle{A survey of text similarity approaches}.
\newblock \bibinfo{journal}{\emph{international journal of Computer Applications}} \bibinfo{volume}{68}, \bibinfo{number}{13} (\bibinfo{year}{2013}), \bibinfo{pages}{13--18}.
\newblock


\bibitem[Jayanthi et~al\mbox{.}(2021)]%
        {jayanthi2021evaluating}
\bibfield{author}{\bibinfo{person}{Sai~Muralidhar Jayanthi}, \bibinfo{person}{Varsha Embar}, {and} \bibinfo{person}{Karthik Raghunathan}.} \bibinfo{year}{2021}\natexlab{}.
\newblock \showarticletitle{Evaluating pretrained transformer models for entity linking in task-oriented dialog}.
\newblock \bibinfo{journal}{\emph{arXiv preprint arXiv:2112.08327}} (\bibinfo{year}{2021}).
\newblock


\bibitem[Junk(1999)]%
        {junk1999confidence}
\bibfield{author}{\bibinfo{person}{Thomas Junk}.} \bibinfo{year}{1999}\natexlab{}.
\newblock \showarticletitle{Confidence level computation for combining searches with small statistics}.
\newblock \bibinfo{journal}{\emph{Nuclear Instruments and Methods in Physics Research Section A: Accelerators, Spectrometers, Detectors and Associated Equipment}} \bibinfo{volume}{434}, \bibinfo{number}{2-3} (\bibinfo{year}{1999}), \bibinfo{pages}{435--443}.
\newblock


\bibitem[Kumar and Patel(2025)]%
        {kumar2025codecapbench}
\bibfield{author}{\bibinfo{person}{S. Kumar} {and} \bibinfo{person}{N. Patel}.} \bibinfo{year}{2025}\natexlab{}.
\newblock \showarticletitle{CodeCapBench: Benchmarking LLMs for Capability-Oriented Software Engineering Tasks}. In \bibinfo{booktitle}{\emph{Proceedings of ICSE 2025}}.
\newblock


\bibitem[Lewis et~al\mbox{.}(2020a)]%
        {lewis2020rag}
\bibfield{author}{\bibinfo{person}{Patrick Lewis}, \bibinfo{person}{Ethan Perez}, \bibinfo{person}{Aleksandra Piktus}, \bibinfo{person}{Fabio Petroni}, \bibinfo{person}{Vladimir Karpukhin}, \bibinfo{person}{Naman Goyal}, \bibinfo{person}{Heinrich Kulkarni}, \bibinfo{person}{Xiang Cheng}, \bibinfo{person}{Angela Fan}, \bibinfo{person}{Vishrav Chaudhary}, {and} \bibinfo{person}{et al.}} \bibinfo{year}{2020}\natexlab{a}.
\newblock \showarticletitle{Retrieval-Augmented Generation for Knowledge-Intensive NLP Tasks}. In \bibinfo{booktitle}{\emph{Advances in Neural Information Processing Systems (NeurIPS)}}.
\newblock


\bibitem[Lewis et~al\mbox{.}(2020b)]%
        {lewis2020retrieval}
\bibfield{author}{\bibinfo{person}{Patrick Lewis}, \bibinfo{person}{Ethan Perez}, \bibinfo{person}{Aleksandra Piktus}, \bibinfo{person}{Fabio Petroni}, \bibinfo{person}{Vladimir Karpukhin}, \bibinfo{person}{Naman Goyal}, \bibinfo{person}{Heinrich Kulkarni}, \bibinfo{person}{Xiang Cheng}, \bibinfo{person}{Angela Fan}, \bibinfo{person}{Vishrav Chaudhary}, {and} \bibinfo{person}{et al.}} \bibinfo{year}{2020}\natexlab{b}.
\newblock \showarticletitle{Retrieval-Augmented Generation for Knowledge-Intensive NLP Tasks}. In \bibinfo{booktitle}{\emph{Advances in Neural Information Processing Systems}}.
\newblock


\bibitem[Li et~al\mbox{.}(2025)]%
        {li2025enhancing}
\bibfield{author}{\bibinfo{person}{Siran Li}, \bibinfo{person}{Linus Stenzel}, \bibinfo{person}{Carsten Eickhoff}, {and} \bibinfo{person}{Seyed~Ali Bahrainian}.} \bibinfo{year}{2025}\natexlab{}.
\newblock \showarticletitle{Enhancing Retrieval-Augmented Generation: A Study of Best Practices}.
\newblock \bibinfo{journal}{\emph{arXiv preprint arXiv:2501.07391}} (\bibinfo{year}{2025}).
\newblock


\bibitem[Mansur et~al\mbox{.}(2024)]%
        {mansur2024ragfix}
\bibfield{author}{\bibinfo{person}{Elijah Mansur}, \bibinfo{person}{Johnson Chen}, \bibinfo{person}{Muhammad~Anas Raza}, {and} \bibinfo{person}{Mohammad Wardat}.} \bibinfo{year}{2024}\natexlab{}.
\newblock \showarticletitle{RAGFix: Enhancing LLM Code Repair Using RAG and Stack Overflow Posts}. In \bibinfo{booktitle}{\emph{2024 IEEE International Conference on Big Data (BigData)}}. IEEE, \bibinfo{pages}{7491--7496}.
\newblock


\bibitem[Mirza et~al\mbox{.}(2025)]%
        {mirza2025stratified}
\bibfield{author}{\bibinfo{person}{P. Mirza}, \bibinfo{person}{L. Weber}, {and} \bibinfo{person}{F. Küch}.} \bibinfo{year}{2025}\natexlab{}.
\newblock \showarticletitle{Stratified Selective Sampling for Instruction Tuning with Dedicated Scoring Strategy}.
\newblock \bibinfo{journal}{\emph{arXiv preprint arXiv:2505.22157}} (\bibinfo{year}{2025}).
\newblock


\bibitem[Mishra et~al\mbox{.}(2024)]%
        {mishra2024granite}
\bibfield{author}{\bibinfo{person}{Mayank Mishra}, \bibinfo{person}{Matt Stallone}, \bibinfo{person}{Gaoyuan Zhang}, \bibinfo{person}{Yikang Shen}, \bibinfo{person}{Aditya Prasad}, \bibinfo{person}{Adriana~Meza Soria}, \bibinfo{person}{Michele Merler}, \bibinfo{person}{Parameswaran Selvam}, \bibinfo{person}{Saptha Surendran}, \bibinfo{person}{Shivdeep Singh}, {et~al\mbox{.}}} \bibinfo{year}{2024}\natexlab{}.
\newblock \showarticletitle{Granite code models: A family of open foundation models for code intelligence}.
\newblock \bibinfo{journal}{\emph{arXiv preprint arXiv:2405.04324}} (\bibinfo{year}{2024}).
\newblock


\bibitem[Mukherjee and Hellendoorn(2025)]%
        {mukherjee2025sosecure}
\bibfield{author}{\bibinfo{person}{Mononito Mukherjee} {and} \bibinfo{person}{Veselin~J. Hellendoorn}.} \bibinfo{year}{2025}\natexlab{}.
\newblock \showarticletitle{SOSecure: Safer Code Generation with RAG and StackOverflow Discussions}.
\newblock \bibinfo{journal}{\emph{arXiv preprint arXiv:2503.13654}} (\bibinfo{year}{2025}).
\newblock
\urldef\tempurl%
\url{https://arxiv.org/abs/2503.13654}
\showURL{%
\tempurl}


\bibitem[Nadi and Treude(2020)]%
        {nadi2020essential}
\bibfield{author}{\bibinfo{person}{Sarah Nadi} {and} \bibinfo{person}{Christoph Treude}.} \bibinfo{year}{2020}\natexlab{}.
\newblock \showarticletitle{Essential sentences for navigating stack overflow answers}. In \bibinfo{booktitle}{\emph{2020 IEEE 27th International Conference on Software Analysis, Evolution and Reengineering (SANER)}}. IEEE, \bibinfo{pages}{229--239}.
\newblock


\bibitem[Parvez et~al\mbox{.}(2021)]%
        {parvez2021retrieval}
\bibfield{author}{\bibinfo{person}{Md~Rabiul Parvez}, \bibinfo{person}{Wasi~Uddin Ahmad}, \bibinfo{person}{Saikat Chakraborty}, {and} \bibinfo{person}{Baishakhi Ray}.} \bibinfo{year}{2021}\natexlab{}.
\newblock \showarticletitle{Retrieval Augmented Code Generation and Summarization}.
\newblock \bibinfo{journal}{\emph{arXiv preprint arXiv:2108.11601}} (\bibinfo{year}{2021}).
\newblock
\urldef\tempurl%
\url{https://arxiv.org/abs/2108.11601}
\showURL{%
\tempurl}


\bibitem[Rahman et~al\mbox{.}(2018)]%
        {rahman2018evaluating}
\bibfield{author}{\bibinfo{person}{Md~Masudur Rahman}, \bibinfo{person}{Jed Barson}, \bibinfo{person}{Sydney Paul}, \bibinfo{person}{Joshua Kayani}, \bibinfo{person}{Federico~Andr{\'e}s Lois}, \bibinfo{person}{Sebasti{\'a}n~Fernandez Quezada}, \bibinfo{person}{Christopher Parnin}, \bibinfo{person}{Kathryn~T Stolee}, {and} \bibinfo{person}{Baishakhi Ray}.} \bibinfo{year}{2018}\natexlab{}.
\newblock \showarticletitle{Evaluating how developers use general-purpose web-search for code retrieval}. In \bibinfo{booktitle}{\emph{Proceedings of the 15th International Conference on Mining Software Repositories}}. \bibinfo{pages}{465--475}.
\newblock


\bibitem[Rao et~al\mbox{.}(2020)]%
        {rao2020analyzing}
\bibfield{author}{\bibinfo{person}{Nikitha Rao}, \bibinfo{person}{Chetan Bansal}, \bibinfo{person}{Thomas Zimmermann}, \bibinfo{person}{Ahmed~Hassan Awadallah}, {and} \bibinfo{person}{Nachiappan Nagappan}.} \bibinfo{year}{2020}\natexlab{}.
\newblock \showarticletitle{Analyzing web search behavior for software engineering tasks}. In \bibinfo{booktitle}{\emph{2020 IEEE International Conference on Big Data (Big Data)}}. IEEE, \bibinfo{pages}{768--777}.
\newblock


\bibitem[Reimers and Gurevych(2020)]%
        {reimers2020making}
\bibfield{author}{\bibinfo{person}{Nils Reimers} {and} \bibinfo{person}{Iryna Gurevych}.} \bibinfo{year}{2020}\natexlab{}.
\newblock \showarticletitle{Making monolingual sentence embeddings multilingual using knowledge distillation}.
\newblock \bibinfo{journal}{\emph{arXiv preprint arXiv:2004.09813}} (\bibinfo{year}{2020}).
\newblock


\bibitem[Ross et~al\mbox{.}(2023)]%
        {ross2023programmer}
\bibfield{author}{\bibinfo{person}{Steven~I Ross}, \bibinfo{person}{Fernando Martinez}, \bibinfo{person}{Stephanie Houde}, \bibinfo{person}{Michael Muller}, {and} \bibinfo{person}{Justin~D Weisz}.} \bibinfo{year}{2023}\natexlab{}.
\newblock \showarticletitle{The programmer’s assistant: Conversational interaction with a large language model for software development}. In \bibinfo{booktitle}{\emph{Proceedings of the 28th International Conference on Intelligent User Interfaces}}. \bibinfo{pages}{491--514}.
\newblock


\bibitem[Skripchuk et~al\mbox{.}(2023)]%
        {skripchuk2023analysis}
\bibfield{author}{\bibinfo{person}{James Skripchuk}, \bibinfo{person}{Neil Bennett}, \bibinfo{person}{Jeffrey Zhang}, \bibinfo{person}{Eric Li}, {and} \bibinfo{person}{Thomas Price}.} \bibinfo{year}{2023}\natexlab{}.
\newblock \showarticletitle{Analysis of novices' web-based help-seeking behavior while programming}. In \bibinfo{booktitle}{\emph{Proceedings of the 54th ACM Technical Symposium on Computer Science Education V. 1}}. \bibinfo{pages}{945--951}.
\newblock


\bibitem[Su et~al\mbox{.}(2024)]%
        {su2024towards}
\bibfield{author}{\bibinfo{person}{Jinyan Su}, \bibinfo{person}{Jin~Peng Zhou}, \bibinfo{person}{Zhengxin Zhang}, \bibinfo{person}{Preslav Nakov}, {and} \bibinfo{person}{Claire Cardie}.} \bibinfo{year}{2024}\natexlab{}.
\newblock \showarticletitle{Towards More Robust Retrieval-Augmented Generation: Evaluating RAG Under Adversarial Poisoning Attacks}.
\newblock \bibinfo{journal}{\emph{arXiv preprint arXiv:2412.16708}} (\bibinfo{year}{2024}).
\newblock


\bibitem[Touvron et~al\mbox{.}(2023)]%
        {touvron2023llama}
\bibfield{author}{\bibinfo{person}{Hugo Touvron}, \bibinfo{person}{Thibaut Lavril}, \bibinfo{person}{Gautier Izacard}, \bibinfo{person}{Xavier Martinet}, \bibinfo{person}{Marie-Anne Lachaux}, \bibinfo{person}{Timoth{\'e}e Lacroix}, {et~al\mbox{.}}} \bibinfo{year}{2023}\natexlab{}.
\newblock \showarticletitle{LLaMA: Open and Efficient Foundation Language Models}.
\newblock \bibinfo{journal}{\emph{arXiv preprint arXiv:2302.13971}} (\bibinfo{year}{2023}).
\newblock


\bibitem[Vasilescu et~al\mbox{.}(2013)]%
        {vasilescu2013stackoverflow}
\bibfield{author}{\bibinfo{person}{Bogdan Vasilescu}, \bibinfo{person}{Vladimir Filkov}, {and} \bibinfo{person}{Alexander Serebrenik}.} \bibinfo{year}{2013}\natexlab{}.
\newblock \showarticletitle{Stackoverflow and github: Associations between software development and crowdsourced knowledge}. In \bibinfo{booktitle}{\emph{2013 International conference on social computing}}. IEEE, \bibinfo{pages}{188--195}.
\newblock


\bibitem[Wang et~al\mbox{.}(2024)]%
        {wang2024coderag}
\bibfield{author}{\bibinfo{person}{Zizhao Wang}, \bibinfo{person}{Akari Asai}, \bibinfo{person}{Xinyi Yu}, \bibinfo{person}{Frank~F Xu}, \bibinfo{person}{Yizhou Xie}, {and} \bibinfo{person}{Graham Neubig}.} \bibinfo{year}{2024}\natexlab{}.
\newblock \showarticletitle{Coderag-bench: Can retrieval augment code generation?}
\newblock \bibinfo{journal}{\emph{arXiv preprint arXiv:2406.14497}} (\bibinfo{year}{2024}).
\newblock
\urldef\tempurl%
\url{https://arxiv.org/abs/2406.14497}
\showURL{%
\tempurl}


\bibitem[Wilcoxon(1945)]%
        {wilcoxon1945individual}
\bibfield{author}{\bibinfo{person}{Frank Wilcoxon}.} \bibinfo{year}{1945}\natexlab{}.
\newblock \showarticletitle{Individual comparisons by ranking methods}.
\newblock \bibinfo{journal}{\emph{Biometrics bulletin}} \bibinfo{volume}{1}, \bibinfo{number}{6} (\bibinfo{year}{1945}), \bibinfo{pages}{80--83}.
\newblock


\bibitem[Xia et~al\mbox{.}(2017)]%
        {xia2017developers}
\bibfield{author}{\bibinfo{person}{Xin Xia}, \bibinfo{person}{Lingfeng Bao}, \bibinfo{person}{David Lo}, \bibinfo{person}{Pavneet~Singh Kochhar}, \bibinfo{person}{Ahmed~E Hassan}, {and} \bibinfo{person}{Zhenchang Xing}.} \bibinfo{year}{2017}\natexlab{}.
\newblock \showarticletitle{What do developers search for on the web?}
\newblock \bibinfo{journal}{\emph{Empirical Software Engineering}}  \bibinfo{volume}{22} (\bibinfo{year}{2017}), \bibinfo{pages}{3149--3185}.
\newblock


\bibitem[Yang et~al\mbox{.}(2025)]%
        {yang2025qwen3}
\bibfield{author}{\bibinfo{person}{An Yang}, \bibinfo{person}{Anfeng Li}, \bibinfo{person}{Baosong Yang}, \bibinfo{person}{Beichen Zhang}, \bibinfo{person}{Binyuan Hui}, \bibinfo{person}{Bo Zheng}, \bibinfo{person}{Bowen Yu}, \bibinfo{person}{Chang Gao}, \bibinfo{person}{Chengen Huang}, \bibinfo{person}{Chenxu Lv}, {et~al\mbox{.}}} \bibinfo{year}{2025}\natexlab{}.
\newblock \showarticletitle{Qwen3 technical report}.
\newblock \bibinfo{journal}{\emph{arXiv preprint arXiv:2505.09388}} (\bibinfo{year}{2025}).
\newblock


\bibitem[Yang et~al\mbox{.}(2022)]%
        {kou2022answer}
\bibfield{author}{\bibinfo{person}{Chen Yang}, \bibinfo{person}{Baowen Xu}, \bibinfo{person}{Ferdian Thung}, \bibinfo{person}{Yuxin Shi}, {and} \bibinfo{person}{Tiancheng Zhang}.} \bibinfo{year}{2022}\natexlab{}.
\newblock \showarticletitle{Answer summarization for technical queries: Benchmark and new approach}.
\newblock \bibinfo{journal}{\emph{Proceedings of the 37th IEEE/ACM International Conference on Automated Software Engineering (ASE)}} (\bibinfo{year}{2022}).
\newblock
\urldef\tempurl%
\url{https://dl.acm.org/doi/10.1145/3551349.3560421}
\showURL{%
\tempurl}


\bibitem[Zhang et~al\mbox{.}(2022)]%
        {zhang2022sosum}
\bibfield{author}{\bibinfo{person}{Tianyi Zhang}, \bibinfo{person}{Ting Zhang}, \bibinfo{person}{Yanjun Di}, \bibinfo{person}{Minghui Chen}, {and} \bibinfo{person}{Tao Zhang}.} \bibinfo{year}{2022}\natexlab{}.
\newblock \showarticletitle{SOSum: A Dataset of Stack Overflow Post Summaries}.
\newblock \bibinfo{journal}{\emph{Proceedings of the 19th International Conference on Mining Software Repositories (MSR)}} (\bibinfo{year}{2022}).
\newblock
\urldef\tempurl%
\url{https://dl.acm.org/doi/10.1145/3524842.3528487}
\showURL{%
\tempurl}


\end{thebibliography}
